\documentclass{article}
\usepackage{icmc2023template}
\usepackage{times}
\usepackage{ifpdf}

\pdfoutput=1 

\usepackage{soul}
\usepackage[english]{babel}

\usepackage{amssymb}

\usepackage{amsmath}

\usepackage{setspace}

\usepackage{subfigure}

\usepackage{multirow}

\def\papertitle{Music Style Transfer With Diffusion Model}
\def\firstauthor{Hong Huang}
\def\secondauthor{Yuyi Wang*} 
\def\thirdauthor{Luyao Li}
\def\fourthauthor{Jun Lin}

\newif\ifpdf
\ifx\pdfoutput\relax
\else
   \ifcase\pdfoutput
      \pdffalse
   \else
      \pdftrue
  \fi
\fi

\ifpdf 
  \usepackage[pdftex,
    pdftitle={\papertitle},
    pdfauthor={\firstauthor, \secondauthor, \thirdauthor, \fourthauthor},
    bookmarksnumbered, 
    pdfstartview=XYZ 
   ]{hyperref}

  \usepackage[pdftex]{graphicx}
  \graphicspath{{./figures/}}
  \DeclareGraphicsExtensions{.pdf,.jpeg,.png}

  \usepackage[figure,table]{hypcap}

\else 
  \usepackage[dvips,
    bookmarksnumbered, 
    pdfstartview=XYZ 
  ]{hyperref}  

  \usepackage[dvips]{epsfig,graphicx}
  \graphicspath{{./figures/}}
  \DeclareGraphicsExtensions{.eps}

  \usepackage[figure,table]{hypcap}
\fi

\hypersetup{
    colorlinks,%
    citecolor=black,%
    filecolor=black,%
    linkcolor=black,%
    urlcolor=black
}

\title{\papertitle}

\fourauthors
  {1.5in}
  {\firstauthor} {Hunan University of Technology, China \\ %
    {\tt \href{mailto:author1@ul.ie}{alex\_hh3344@163.com}}}
  {\secondauthor \thanks{*Corresponding author}} {CRRC Zhuzhou Institute, China  \\ %
    {\tt \href{mailto:author2@ul.ie}{yuyiwang920@gmail.com}}}
  {\thirdauthor} { Hunan University of Technology, China \\ %
    {\tt \href{mailto:author3@ul.ie}{l1753554929@163.com}}}
  {\fourthauthor} { CRRC Zhuzhou Institute, China \\ %
    {\tt \href{mailto:author4@ul.ie}{linj@163.com}}}

\begin{document}
\capstartfalse
\maketitle
\capstarttrue
\begin{abstract} 
Previous studies on music style transfer have mainly focused on one-to-one style conversion, which is relatively limited. When considering the conversion between multiple styles, previous methods required designing multiple modes to disentangle the complex style of the music, resulting in large computational costs and slow audio generation. The existing music style transfer methods generate spectrograms with artifacts, leading to significant noise in the generated audio. To address these issues, this study proposes a music style transfer framework based on diffusion models (DM) and uses spectrogram-based methods to achieve multi-to-multi music style transfer. The GuideDiff method is used to restore spectrograms to high-fidelity audio, accelerating audio generation speed and reducing noise in the generated audio. Experimental results show that our model has good performance in multi-mode music style transfer compared to the baseline and can generate high-quality audio in real-time on consumer-grade GPUs.
\end{abstract}

\section{Introduction}\label{sec:introduction}
The study of musical styles is important for the development of music. Incorporating different styles into compositions can lead to new and innovative music. Transferring musical styles can create works that pay homage to traditional styles while incorporating contemporary elements. By studying how different styles can be combined and transformed, musicians can create new forms of artistic expression.

When discussing the transfer of musical style, it is typically believed that music can be broken down into two elements: content and style. The goal of music style transfer is to maintain the content of the music while modifying the style. With the rapid development of deep generative models, various models such as autoregressive models, generative adversarial networks, variational autoencoders, and stream-based models have actively promoted the development of speech synthesis and music generation.
Furthermore, many academics have used these models to research musical style transfer. MIDI-VAE, a neural network model based on variational autoencoders, was used by Brunner et al. \cite{brunner2018midi} to convert the style of polyphonic music with several instrumental tracks.
The same year, Brunner et al. \cite{brunner2018symbolic} offered a different approach that involved converting midi format audio into a piano rolling matrix, training CycleGAN with the matrix, and then producing converted midi audio. However, this method can only transfer the style from the playing dimension.
Huang et al. \cite{huang2018timbretron} proposed Timbretron by extracting CQT features of the audio, then converting them into timbre through CycleGAN, and finally synthesizing CQT features into original audio waveforms using pre-trained WaveNet. Their method can capture higher resolution at lower frequencies and maintain equal variance of pitch energy, but the generated audio quality is still inadequate.
Donahue et al. \cite{donahue2019lakhnes} enhanced the effect of multi-instrument music generation through cross-domain training based on Transformer, but the quality of synthesized audio is still inadequate. Hung et al. \cite{hung2019musical} proposed a deep learning model for rearranging any music, resulting in a "stylistic shift" without much impact on the tonal substance.
Bonnici et al. \cite{bonnici2022timbre} used a variational autoencoder combined with a generative adversarial network to construct a meaningful representation of source audio and generate a realistic generation of the target audio.
Noam et al. \cite{mor2018universal} proposed a general music translation network that achieves timbre conversion by training a WaveNet encoder and multiple WaveNet decoders. This method can convert from one timbre domain to multiple timbre domains, but it requires training multiple decoders to adapt to different styles, which is computationally expensive, and the synthesized audio speed is slow.
Denoising Diffusion Probability Models (DDPMs) \cite{ho2020denoising} and Score Matching (SM) \cite{song2020score} are recently proposed methods that have achieved good results in the fields of speech synthesis and music generation. The aforementioned studies have achieved promising results in their respective research directions, but they mainly focus on transferring a single attribute of music (timbre, performance style, composition style), and previous methods suffer from artifacts in the generated spectrograms. Considering many-to-many style migration, previous methods have suffered from complex design structures, high computational overhead, and slow generation of audio.\\
\indent\setlength{\parindent}{2mm}To overcome these limitations, this study uses DM, another type of generative model, whose synthesis process extracts the required generated samples from noise through iterative steps. As the number of iterations increases, the quality of the synthesis improves. However, directly extending DMs to audio generation requires a large amount of computational resources \cite{mittal2021symbolic} and cannot solve the problem of slow generation speed. To address these issues, this study proposes a general and efficient music style transfer framework based on the latent diffusion model (LDM) \cite{rombach2022high}. Specifically, the framework consists of two parts: style transfer and audio generation. In the style transfer part, a conditional mechanism is introduced to learn different types of input styles and transfer their information to the latent space for guiding the generation of target spectrograms. This approach avoids the need for designing complex, disentangled transfer frameworks and enables many-to-many style transfer. Moreover, the transfer process takes place in latent space, greatly reducing computational costs and improving generation speed. For the audio generation part, this study proposes GuideDiff, a waveform audio generator based on DMs. It compresses and encodes spectrograms into the latent space to control and guide waveform generation, achieving fast inference speed and high-quality audio generation compared to baseline vocoders. This has practical significance for the real-time generation of high-quality audio.\\ 
\indent\setlength{\parindent}{2mm}In summary, the main works are as follows:\\ 
\indent\setlength{\parindent}{2mm}
(1)	The paper introduces a music style transfer model that is based on DM and allows for many-to-many music styles to be transferred. This model is capable of performing real-time style transfer on audio, making it highly efficient and practical.\\ 
\indent\setlength{\parindent}{2mm}
(2) This study proposes a novel audio generation method called GuideDiff, which is based on the diffusion model. The GuideDiff method is designed to generate high-quality audio waveforms by utilizing spectrogram restoration techniques. \\
\indent\setlength{\parindent}{2mm}
(3) The experimental results show that the proposed model has good performance in both style transfer and audio quality compared to the baseline model. Moreover, it can achieve real-time conversion and generate target audio on consumer-grade GPUs. \\
\indent\setlength{\parindent}{2mm}
In the remainder of this paper, we will organize the content as follows: Section 2 presents related work; Section 3 describes the architecture of the proposed method; Section 4 evaluates the effectiveness of the proposed method through experiments; and Section 5 provides the conclusion of this paper.

\section{Related Work}

\subsection{Music Style Transfer}
Numerous studies on musical style transfer have taken cues from models for transferring image styles. Musical style transfer can be categorized into three types: timbral style transfer, performance style transfer, and compositional style transfer. Among these, timbral-style transfer has received the most attention in recent years. This type of transfer involves altering the timbre of a musical composition in the audio domain. However, relatively little study has been done on the latter two types of musical style transfer: performance and compositional. Further study on these types of musical style transfers could lead to new and innovative ways of creating and transforming music.

Researchers typically follow two different design patterns to achieve music style transfer. One involves symbolic music notation, and the other involves audio signals. For audio signals, researchers typically use time-frequency methods, which are more indirect and help reduce data complexity. They convert the abstract audio into spectrograms and use deep learning models for high-quality transfer. This method involves two deep learning models, with the first model involving the style transfer of the spectrogram of the audio and the second model involving the restoration of the generated spectrogram to real audio. Currently, researchers mainly use generative models such as CycleGAN \cite{brunner2018symbolic}, VAE \cite{brunner2018midi}, UNIT \cite{liu2017unsupervised}, and MusicVAE \cite{roberts2018hierarchical} for music style transfer. However, while these models have shown promising results, they also have limitations that hinder their practical application. Further research is needed to overcome these limitations and improve the effectiveness and efficiency of musical style transfer.

The focus of this study is to explore a new generic music style transfer model that employs a time-frequency approach. This model is designed to enable three types of music style transfer: timbral, performance, and compositional.

\subsection{Diffusion Models}
DM is a class of likelihood-based generative models, with its pioneering work being DDPM. Its core theoretical underpinnings are the Markov chain and Langevin dynamics. Due to its stable training and easy expansion, it has surpassed GANs \cite{dhariwal2021diffusion} in image generation tasks and achieved higher sample quality. However, the sampling process is slow, and it needs to follow a Markov chain to generate a sample step by step. DDIM \cite{song2020denoising} accelerates the sampling process by iterative non-Markovian methods while keeping the training process unchanged. ADM \cite{dhariwal2021diffusion} ultimately outperforms GAN-based methods through a well-designed architecture and classifier guidance. A latent diffusion model \cite{rombach2022high} has also been proposed recently for image synthesis. This model compresses the image from pixel space to latent space for diffusion, resulting in significantly reduced computational complexity while achieving high-quality image generation. However, the application of this model in the field of music generation has not been extensively studied. In this study, we propose a generic music style transfer framework based on the latent diffusion model, using spectrograms as an intermediate representation of music. In this respect, our work has something in common with riffusion \cite{riffusion2022}, as both utilize Fourier transforms to process audio waveforms in order to obtain a spectrogram. This spectrogram is then diffused using a diffusion model.

\subsection{Neural Vocoder}
Deep generative models have achieved significant success in modeling audio generation, with common methods including autoregressive models, flow-based models, and diffusion models. WaveNet \cite{oord2016wavenet} is an autoregressive model that generates high-fidelity audio, but its synthesis is slow, and the synthesized audio contains audible noise. WaveRNN \cite{gritsenko2020spectral} is another autoregressive model that reduces computational complexity by using sparse recurrent neural networks. Stream-based models, such as WaveFlow \cite{ping2020waveflow}, WaveGlow \cite{prenger2019waveglow}, and FloWaveNet \cite{kim2018flowavenet}, improve the quality of audio synthesis by maximizing the likelihood of training the model. Recently, DM-based audio generation models have been proposed, such as DiffWave \cite{kong2020diffwave} and WaveGrad \cite{chen2020wavegrad}, which are able to generate higher-quality audio and synthesize it faster than common models.

In this work, we propose a new neural vocoder called GuideDiff based on DM. This vocoder is mainly used in the style transfer model to restore high-quality audio from generated spectrograms. Moreover, its synthesis speed is several orders of magnitude faster than baseline models like WaveNet.

\section{Method}

\begin{figure}[h]
\centering
\includegraphics[width=0.9\columnwidth]{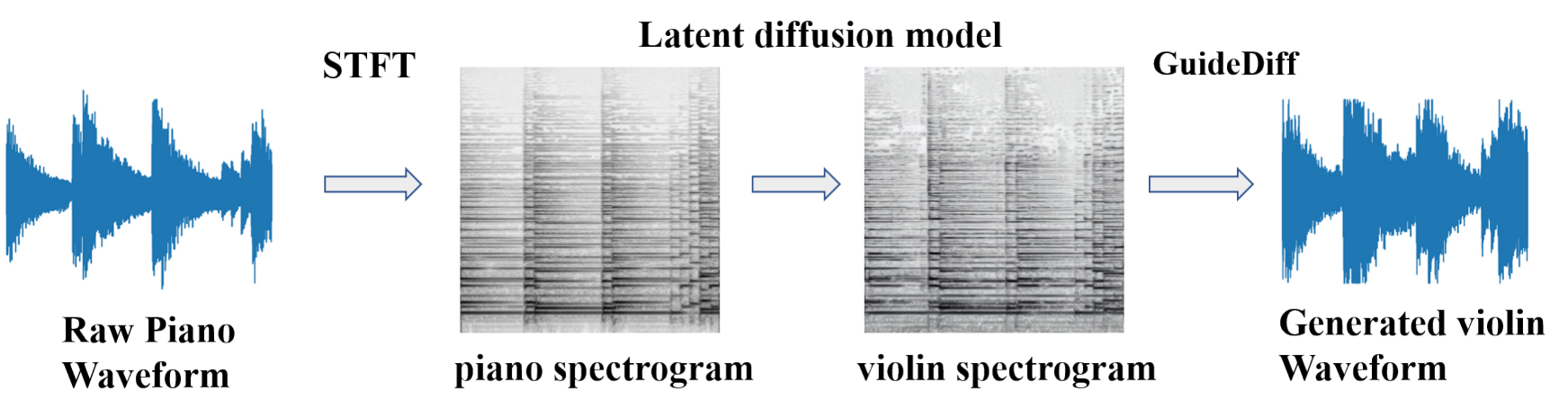}
\caption{Piano to violin style transfer.\label{fig:1}}
\end{figure}
Music style transfer is accomplished in three steps in this work, as illustrated in Figure \ref{fig:1}. Firstly, a spectrogram is obtained from the input audio waveform using the Short Time Fourier Transform (STFT), which represents time and frequency. The phase information is discarded, and only the amplitude is processed as an image. Secondly, the transfer of musical styles is performed by completing the domain conversion on the spectrogram using a latent diffusion model. Lastly, GuideDiff is utilized to convert the transformed spectrograms into audio waveforms.

The section following this introduction will focus on the second part of the music style transfer process, which involves the conversion of the input spectrogram to a target spectrogram using a latent diffusion model. The subsequent section will cover the third part of the process, which is the conversion of the target spectrogram into a high-quality audio waveform using the proposed neural coder, GuideDiff.

\subsection{Time-Frequency Analysis}
\begin{figure}[h]
\centering
\includegraphics[width=0.5\columnwidth]{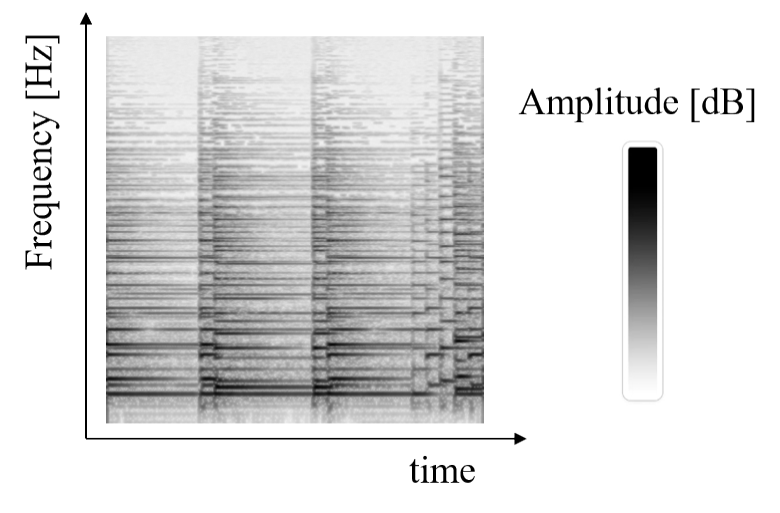}
\caption{Spectrogram.\label{fig:2}}
\end{figure}

The audio signal is often more challenging to capture compared to image signals. As a result, an audio spectrogram, which provides a visual representation of the frequency content of sound, is commonly used. In Figure \ref{fig:2}, the x-axis represents time, while the y-axis represents frequency. The color of each pixel corresponds to the frequency and volume of the audio in its corresponding rows and columns. To perform style transfer, we need to analyze the input audio in both the time and frequency domains to obtain a spectrogram.

One of the most commonly used techniques in this area is the Short Time Fourier Transform (STFT), which is often discretized for computer calculations. The discrete STFT operation can be abbreviated as
\begin{equation}
STFT{x[n]}(m,\omega _{k} )=\sum_{n=-\infty }^{\infty } x[n]\omega [n-m]e^{-j\omega _{k}n } 
\label{eq:1}
\end{equation}
Where $x[n]$ is the input time domain signal, $m$ is the step size, $\omega _{k}$ is the frequency, and $\omega$ is a window function.

The audio is divided into segments of 5 seconds for time-frequency analysis to make processing easier. By performing the STFT transform independently, the segmented audio is converted into a spectrogram. In this case, a Hanning window with a step size of 100 is used, and the phase information is discarded during processing because it is ambiguous and unpredictable.


\subsection{Transfer Model}
\begin{figure}[h]
\centering
\includegraphics[width=0.9\columnwidth]{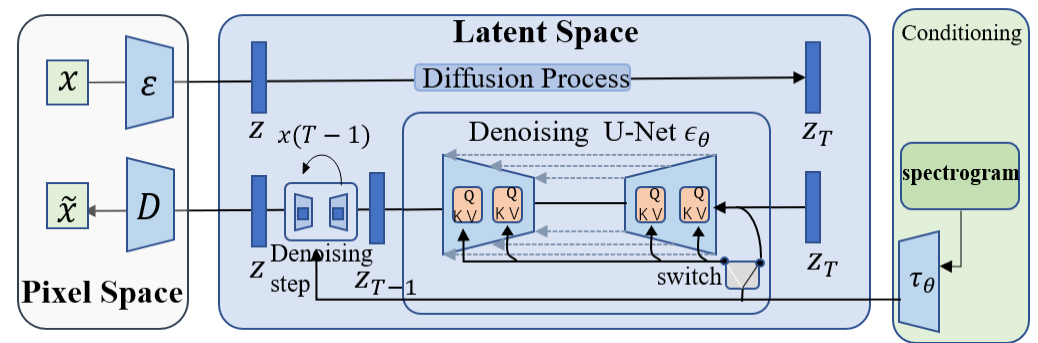}
\caption{Models of transfer.\label{fig:3}}
\end{figure}
Figure \ref{fig:3} illustrates the three main components of the style transfer model: an autoencoder (AE) that compresses and restores the input and output spectrogram information in pixel space; a latent space diffusion model that is mainly used for style transfer, which incorporates a cross-attention mechanism that completes the domain transformation by transferring data from the conditional mechanism into the denoised UNet; the conditional mechanism is primarily used to convey information learned from various musical spectrograms into latent space.

\subsubsection{Perceptual Compression} 
By drawing on the work of Robin Rombach et al. \cite{rombach2022high}, we introduced perceptual compression to lower the computing needs of training DM for producing high-quality spectrograms. The sampling is carried out in a low-dimensional space, which increases the DM's computing efficiency.

A pre-trained self-encoder was employed for perceptual compression. This self-encoder is trained using a patch-based adversarial objective in conjunction with a perceptual loss. The blur created by relying simply on pixel space loss is effectively avoided, which enhances the reconstruction's realism. It offers a low-dimensional representation space that is analogous to the data space from a perceptual standpoint.

The self-encoder consists of an encoder $\varepsilon$ and a generator $D$. They are both composed of three layers of three-dimensional convolution. Formally, given a sample spectrogram $x\in \mathbb{R} ^{H\times W\times 3} $, the encoder $\varepsilon$ encodes it into a potential representation $z=\varepsilon (x)$,where $z\in \mathbb{R} ^{h\times w\times3}$. The encoder downsamples the spectrogram by a factor $f=H/h=W/w$ and the generator $D$ reconstructs the potential representation back into a sample $\tilde{x} $,i.e. $\tilde{x} =D(z)$.

To avoid a high degree of dissimilarity in the potential representation space, we have adopted a KL-reg regularization, introducing a slight KL penalty term. A standard learning rate is obtained at the beginning, and the effect is very close to that of a variational autoencoder (VAE). The reconstruction loss $\mathcal{L} _{rec} $ consists of pixel-level mean squared error (MSE) and perceptual-level loss. In summary, the overall training objectives for encoder $\varepsilon$ and generator $D$ are
\begin{equation}
\mathcal{L} _{AE} =\min_{\varepsilon,D } (\mathcal{L} _{rec}(x,D(\varepsilon(x)))+KL_{reg}(x||(\varepsilon (x)))) 
\label{eq:2}
\end{equation}

\subsubsection{Latent Diffusion Models}
With the perceptual compression model, we can obtain an effective, low-dimensional latent space in which high frequencies and some difficult-to-perceive details are abstracted. This is effective for the extraction of musical features such as pitch, loudness, timbre, etc. Recalling DM, we propose to diffuse and denoise the spectrogram in the latent space. Given a compressed latent code $z_{0}\sim  q(z_{0})$. DM consists of a forward diffusion process and a backward denoising process. In the forward diffusion process, we train the diffusion model by iteratively adding $T$ steps of diffusion Gaussian noise according to a fixed noise schedule, starting from data $z_{0}$ to produce a set of noisy latent variables, i.e. $z_{1} ,...,z_{T}$ .
\begin{equation}
q(z_{t}|z_{t-1}  )=\mathcal{N} (z_{t};\sqrt{1-\beta }_{t}z_{t-1}, \beta _{t}I  )
\label{eq:3}
\end{equation}
\begin{equation}
q(z_{1:T}|z_{0}  )=\prod_{t=1}^{T} q(z_{t}|z_{t-1}  )
\label{eq:4}
\end{equation}
where $\beta _{1} ,\beta _{2},...,\beta  _{T} $ is the noise scheduling that converts the data distribution $z_{0}$ into a potential $z_{T}$.

Ultimately, data points $z_{T}$ are indistinguishable from pure Gaussian noise when mixed together. The diffusion model is employed in the reverse denoising process to recover $z_{T}$ to $z_{0}$ by
\begin{equation}
p(z_{t-1} |z_{t})=\mathcal{N} (z_{t-1};\mu_{\theta } (z_{t} ,t) ,\sigma_{\theta } (z_{t},t))
\label{eq:5}
\end{equation}
\begin{equation}
p_{\theta }(z_{0:T})=p(z_{T})\prod_{t=1}^{T}p_{\theta }(z_{t-1}|z_{t}) 
\label{eq:6}
\end{equation}
where $\theta$ is a parameterized neural network that is defined by a Markov chain. The U-Net, commonly used in image synthesis, is used here to predict $\mu_{\theta } (z_{t} ,t)$ and $\sigma_{\theta } (z_{t},t)$. In actuality, $\sigma_{\theta }$ is set to a time-dependent constant that is untrained depending on a noise schedule of
\begin{equation}
\sigma _{\theta }(z_{t},t)=\sigma _{t}=\frac{1-\bar{\alpha }_{t-1} }{1-\bar{\alpha }_{t}}\beta _{t}  
\label{eq:7}
\end{equation}
Where $\alpha _{t}=1-\beta _{t}$,$\bar{a} _{t}=\prod_{i=1}^{t} \alpha _{i}$, we parameterize $\mu _{\theta }=(z_{t},t)$ by
\begin{equation}
\mu _{\theta }(z_{t},t)=\frac{1}{\sqrt{\alpha _{t}} } (z_{t}-\frac{\beta _{t}}{\sqrt{1-\bar{\alpha }_{t} } }\epsilon _{\theta }(z_{t},t) )  
\label{eq:8}
\end{equation}
Final $\epsilon _{\theta }(z_{t},t)$ was assessed.

In practice, we use simplified training objectives.
\begin{equation}
\mathcal{L}_{simple}(\theta ) =\mathbb{E} _{\varepsilon(x),\epsilon \sim \mathcal{N}(0,1)  }||\epsilon _{\theta }(z_{t},t)-\epsilon ||_{2}^{2}   
\label{eq:9}
\end{equation}
where $\epsilon \sim \mathcal{N}(0,1) $ . Since the forward process of the diffusion model is fixed, it can be efficiently obtained during training $z_{t}$ and the $p(z)$ samples generated by the reverse process can be encoded once in perceptual space through the generator $D$ into image space.

Style transfer module. To model the generation of spectrograms in the latent space and to accomplish style transfer. We used a 2-dimensional convolutional layer to build the underlying U-Net capabilities, specifically a $2\times 2$ shaped convolutional layer. A cross-attention mechanism is added to augment the U-Net backbone, enabling it to generate spectrograms in the target domain conditional on the style transfer. And it ensures that style information can be shared across the potential space, which is essential for learning the style of audio and completing style transfer.

\subsubsection{Conditioning Mechanisms} 
In this module, we employ a domain-specific encoder $\tau _{\theta }$ to preprocess the input conditional style spectrogram $y$ and project the encoded $y$ onto an intermediate representation $\tau _{\theta }(y)\in \mathbb{R} ^{M\times d_{\tau }} $, which is then mapped to the intermediate layer of the U-Net via a cross-attention layer to enable the generation of the spectrograms according to condition $y$. The following equation carries out the cross-attention mechanism.
\begin{equation}
Attention(Q,K,V)=softmax(\frac{QK^{T} }{\sqrt{d} } )\cdot V  
\label{eq:10}
\end{equation}
where $Q=W_{Q}^{(i)} \cdot \varphi _{i}(z_{t})$ ,$K=W_{K}^{(i)} \cdot \tau _{\theta }(y)$,$V=W_{V}^{(i)} \cdot \tau _{\theta }(y)$. $\varphi _{i}(z_{t})\in \mathbb{R} ^{d\times d_{\epsilon }^{i}  } $ denotes the intermediate U-Net representation that implements $\varepsilon _{\theta }$. $W_{V}^{(i)}\in \mathbb{R} ^{d\times d_{\epsilon }^{i}  } 
$,$W_{K}^{(i)}\in \mathbb{R} ^{d\times d_{\epsilon }^{i}  } $ and $W_{Q}^{(i)}\in \mathbb{R} ^{d\times d_{\epsilon }^{i}  } $ are projection matrices that are mainly used to learn and map styles from the target domain of the $\tau _{\theta }(y)$ representation, enabling style transfer. The objective function is rewritten as
\begin{equation}
\mathcal{L} _{CM}(\theta)=\mathbb{E}_{\varepsilon(x),y,\epsilon\sim \mathcal{N}(0,1)}||\epsilon -\epsilon _{\theta }(z_{t},t,\tau _{\theta }(y))||_{2}^{2}     
\label{eq:11}
\end{equation}
Where $\tau _{\theta }$ and $\epsilon _{\theta }$ can be jointly optimized by means of an objective function.

\subsection{Waveform Reconstruction} 
\begin{figure}[h]
\centering
\includegraphics[width=0.9\columnwidth]{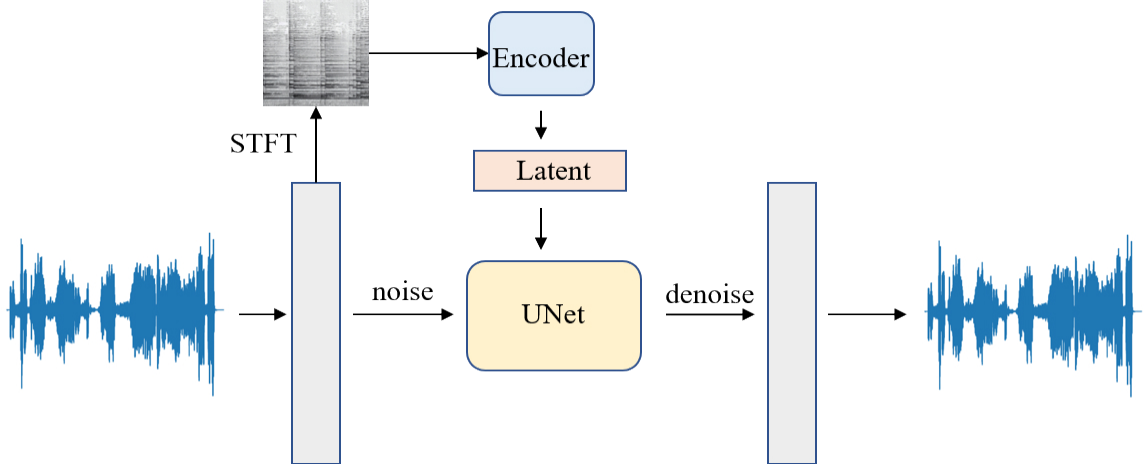}
\caption{GuideDiff architecture.\label{fig:4}}
\end{figure}
\begin{figure}[h]
\centering
\includegraphics[width=0.8\columnwidth]{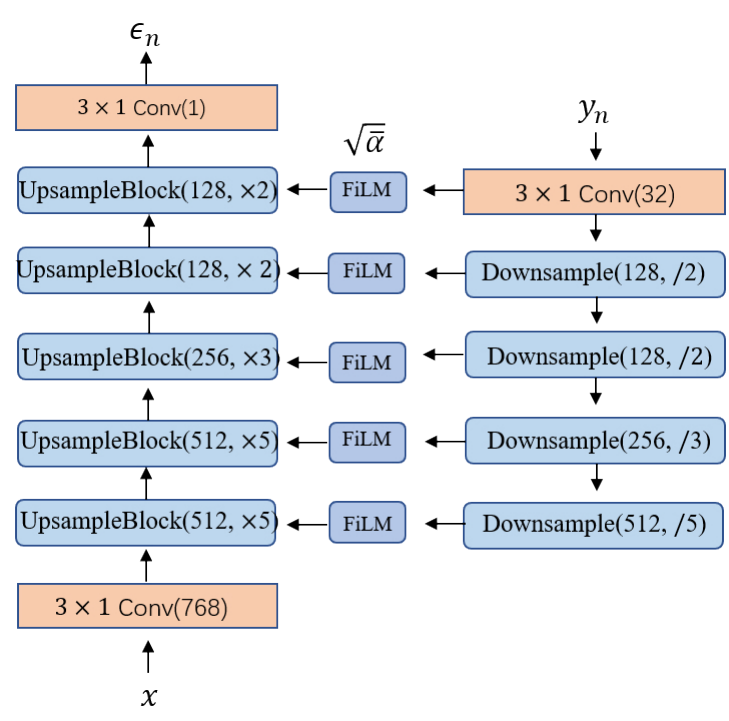}
\caption{Model's primary network.\label{fig:5}}
\end{figure}

We propose a novel encoder, called GuideDiff, to convert the spectrogram output from the model into audio. It can restore the spectrogram to generate high-quality audio. As shown in Figure \ref{fig:4}, a $3\times  3$ encoder $\varepsilon =E_{\theta _{enc}}(m_{w})$ is first used to encode the spectrogram into the latent space, and then the information $x$ from the latent space is sent as conditional information into the U-Net's cross-attention mechanism for conditioning and directing the creation of waveforms. The original waveform is then recreated by using the diffusion decoder $D =D_{\theta _{dec}}(z,\alpha ,s)$ to decode the latent signal, where $D_{\theta _{dec}}$ denotes the diffusion sampling method, $\alpha$ denotes the noise, and $s$ denotes the sample pace length. Target diffusion is used to train Decoder $D$ while conditioning the latent 2D U-Net, which is repeatedly invoked during the decoding procedure. Figure \ref{fig:5} displays the model's primary network diagram.
Where $y_{n}$ denotes the $n$ th round of noisy audio input and $\epsilon _{\theta }$ denotes the simulated generated noise. FiLM is the characteristic linear modulation module, consisting of two $3 \times 1$ convolutional layers and the Leaky ReLU function. Here we condition on noise level $\sqrt{\bar{\alpha} } $ and pass it to the position encoding function.

Compared to the DM objective function, we can write the objective function as
\begin{equation}
\mathcal{L} _{GuideDiff}(\theta )=\mathbb{E} _{\bar{\alpha },\varepsilon  }[||\epsilon -\epsilon _{\theta }(\sqrt{\bar{\alpha }_{n} }y_{0}+\sqrt{1-\bar{\alpha }_{n}  }\epsilon ,x,\sqrt{\bar{\alpha } }   )||_{1}]     
\label{eq:12}
\end{equation}
where $\alpha =1-\beta _{n} $,$\bar{\alpha } _{n}=\sum_{s=1}^{n} \alpha _{s}$ , in this case $\beta _{n}$, is an equivariant sequence from 0 to 1.

For the input spectrograms, we discarded the phase and used only the amplitude. By encoding the spectrograms in latent space, the computational load for the representation can be effectively reduced. Moreover, it enables the diffusion model to learn how to generate waveforms with true phase. The latent space obtained is used as the starting point for the next diffusion phase. The advantage of this is that our model only needs to be trained once. The latent trajectory space also allows for a large number of inference procedures to be performed without requiring retraining. Specifically, once this model is trained, it is only necessary to use a different number of iterations $N$ in the inference process to determine the quality of the computational output. This is useful for rapidly bootstrapping the generation of high-quality raw audio. To ensure that the reduced latent space is available for latent diffusion, we apply the tanh function to the bottleneck, ensuring that the values remain within the range [-1, 1].

In summary, our overall objective function is
\begin{equation}
\mathcal{L} =\mathcal{L}_{AE}+\mathcal{L}_{simple}(\theta )+\mathcal{L}_{CM}(\theta )+\mathcal{L}_{GuideDiff}(\theta ) 
\label{eq:13}
\end{equation}

\section{Experiments}
In Section 4.1, we provide details on the experimental setup, including data description and pre-processing, as well as evaluation metrics. Section 4.2 then provides a detailed description of the implementation.

\subsection{Experimental Setup}

\subsubsection{Data Description And Preprocessing}
The model consumes a large amount of memory when generating an entire song at once. To mitigate this issue, we employed the Demucs model to separate the music into its constituent sources, such as vocals, bass, and drums. Furthermore, each song was divided into smaller segments, which were modeled individually and then reassembled. However, rearranging the segments was challenging, as they differed in downbeat, key, and pace. To address this, we smoothly interpolated cues and seeds in the model's latent space. In a diffusion model, the latent space is a feature vector that encompasses the entire space of possibilities that can be generated by the model. Items that are similar to each other are approached in the latent space, and each value in the latent space is decoded into a feasible output. This makes the audio sound natural.

We require various types of music data to train our model to achieve music style migration. For all the experiments in this study, we used music datasets from multiple source domains collected from the web. This dataset includes over 100,000 WAV audio files of various instruments, genres, and compositional styles. The main instruments include piano, violin, guitar, and others, while the genres mainly consist of jazz, classical, and pop. The data was used for training (80\%), testing (10\%), and validation (10\%).

\subsubsection{Evalutaion Metrics}
The following measures were used to evaluate and analyze the model's performance:

\textbf{Fréchet Audio Distance (FAD)} \cite{kilgour2018fr} The FAD calculates the Fréchet distance between the output generated audio samples and the real audio samples. The smaller the distance between the two data distributions, the more realistic the generated samples will be, which gives a reliable assessment of the difference between them.

\textbf{Accuracy} In this research, five independent style assessment classifiers were trained in order to test the efficacy of the model style transfer. The percentage of styles accurately predicted in each song bar served as a measure of the classifiers' accuracy.

\textbf{Mean Opinion Score(MOS)} In this work, a 5-scale mean opinion score is used to evaluate the proposed model. Where the MOS value suggests that a higher value is preferable. Subjects were asked to rate each of the three questions for each transfer version on a scale of 1 to 5.

\begin{enumerate}
    \item Success in style transfer (ST): whether the target domain is migrated in the generated audio after transfer compared to the original audio.
    \item Content preservation (CT): the extent to which the migrated-generated audio matches the original audio content.
    \item Sound quality (SQ): the generated audio has high or poor sound quality.
\end{enumerate}
A mean score will be used when comparing it to other baseline models. GuideDiff simply evaluates the quality of the generated sound.

\textbf{Inception Score (IS) \cite{salimans2016improved}} To evaluate the level of diversity and quality of the sample generation. IS is an evaluation metric employing a ResNeXT classifier \cite{xie2017aggregated} trained on our dataset and a 10-dimensional logit based on a 1024-dimensional feature vector. To assess the effectiveness of the proposed audio generation model. IS is calculated as
\begin{equation}
IS=exp(E_{x\sim p_{gen}}KL(P\mathcal{F}(x)  ||E_{x^{'}\sim p_{gen}}P\mathcal{F}(x^{'})))
\label{eq:14}
\end{equation}
Where $P\mathcal{F}(x)$ is a multinomial distribution and $E_{x^{'}\sim p_{gen}}P\mathcal{F}(x^{'})$ is an edge label distribution.

\subsubsection{Implementation Details}
This work uses a UNet architecture consisting of 14 layers of stacked convolution blocks and attention blocks as a combination of upsampling and downsampling for the diffusion model, which is based on the work of Robin Rombach et al \cite{rombach2022high}. For the downsampling factor, a downsampling factor of 4 was used. The same hidden size and skip connection layers were set between the layers in the UNet model. The first six layers of the UNet model use $512 \times 512$ input and output channels, followed by two $256 \times256$ and $128 x 128$ input and output channels, respectively. After that, the input and output channels are halved layer by layer. The attention mechanism is used in this work at $16\times16$,$ 8\times8$ and $4\times4$ resolutions. A ResBlock is also added to the UNet module, which receives two inputs: the image x and the embedding corresponding to the timestep. Two linear layers and the time\_emb layer make up the time-step embedding layer. Our compression ratio for the latent space is 64. The audio samples were sampled at a frequency of 16000 Hz, with a channel size of 2, and an amplitude of -10 dB. The model was trained using the Adam optimizer with 500k steps, a learning rate of 5e-5, and a batch size of 100. The batch size for GuideDiff was set to 256. Approximately 1M steps were trained using the Adam optimizer.

The experiments in this research generated audio in less than 5 seconds, which can be regarded as real-time generation, and were trained on 3 NVIDA RTX3090Ti, a GPU capable of running 50 stable diffusion steps.


\subsection{Experimental Analysis}
Four primary musical style transfer tasks were taken into consideration in the trials:
\begin{enumerate}
    \item Stylistic transfer of instrument timbres. Mainly consider piano to guitar (p2g) and piano to violin transitions (p2v). Each transition will do a bilateral transformation.
    \item transfer of musical genres. Genre conversions from jazz to pop (j2p) and jazz to class (j2c) are mainly considered. Each transition will do a bilateral transformation.
    \item Music composition style conversion. Beethoven to Chopin (B2C), and Chopin to Beethoven (C2B) conversions are mainly considered.
    \item Many-to-many style conversions. Conversion of classical piano pieces played mainly by Beethoven to jazz violin in the Chopin style (Bcp2Cjv).
\end{enumerate}
\begin{figure}[h]
\centering
\includegraphics[width=0.7\columnwidth]{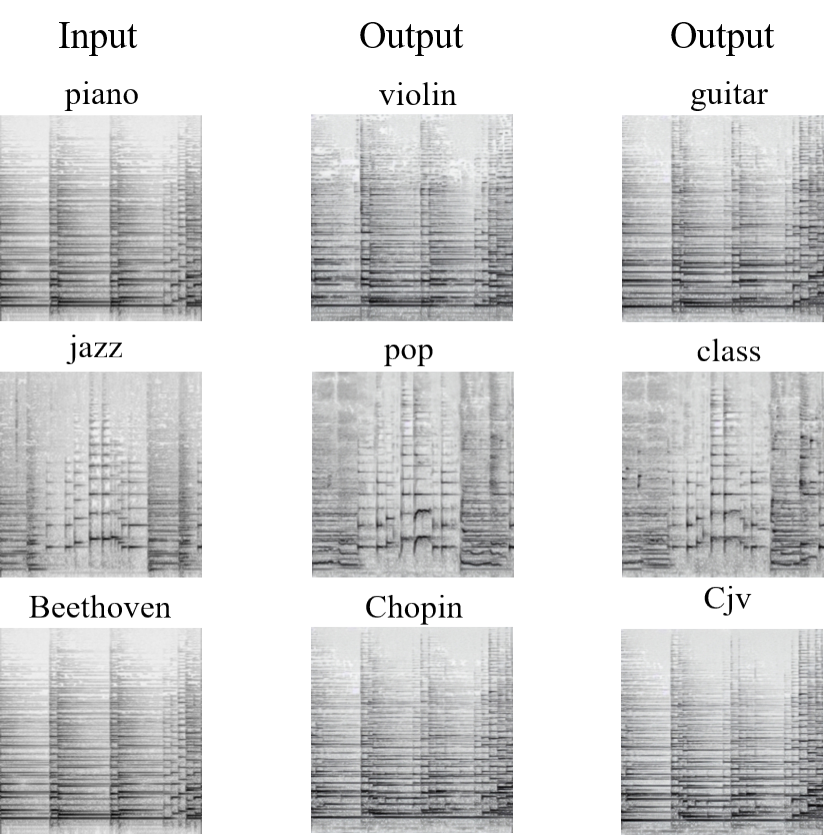}
\caption{Style transitions for various tasks.\label{fig:6}}
\end{figure}

The spectrograms of our model's inputs and outputs for the various tasks are shown in Figure \ref{fig:6}. From the plots, it is clear that the target domain is shifted while the contents are kept intact.

\subsubsection{Style Conversion Evaluation}
This study evaluates the proposed model through four different style transfer challenges. Subjective (MOS) and objective (FAD, accuracy) evaluations are used to compare the style conversions. Each score has its own limitations. Subjective measurements evaluate three main aspects of the model: the success of style transfer (ST), content preservation (CP), and sound quality (SQ). A 5-point scale is used for evaluation. Objective evaluations use FAD to measure individual aspects of the conversion, and accuracy is used to evaluate the accuracy of style transfer.

\textbf{Subjective evaluation} Mean opinion scores (MOS) were collected from 200 testers for the listening test. These testers included both music lovers and non-musicians. In each mid-round score, testers first listened to the original audio clip and then to the style-shifted version.
\begin{table}[h]
 \begin{center}
 \begin{tabular}{l l l l}
  \hline
Task & ST & CP & SQ \\
  \hline  
  piano2Violin & 4.27 & 4.13 & 4.3  \\
  piano2guitar & 4.02 & 4.05 & 4.2 \\
  jazz2pop & 3.95 & 3.8 & 4.0 \\
  jazz2class & 3.96 & 4.0 & 4.12 \\
  Beethoven2Chopin & 4.05 & 4.1 & 4.15 \\  
  Bcp2Cjv & 4.1 & 4.23 & 4.3\\
  \hline
 \end{tabular}
\end{center}
 \caption{ 5-scale MOS with style Transfer.}
 \label{tab:1}
\end{table}

The results indicate that our model performs the best in the piano2violin task, which may be attributed to the relatively simple timbre conversion of a single instrument. Our model's performance is slightly lower in the jazz2pop and jazz2class tasks, but it still achieves scores close to 4 in terms of successful style transfer and content retention. This suggests that our model is relatively successful in genre conversion. Additionally, the high scores for sound quality in all six tasks indicate that the proposed model is capable of generating high-quality music.

\textbf{Objective review} Measures how well the converted version matches the original version and the accuracy of the style transfer.
\begin{table}[h]
 \begin{center}
 \begin{tabular}{l c c }
  \hline
   Task &	FAD$\downarrow$ &	Accuracy$\uparrow$ \\
  \hline
  piano2Violin &	7.52 &	94.5\% \\
  piano2guitar &	6.95 &	93.4\% \\
  jazz2pop &	11.76 &	86.2\% \\
  jazz2class &	10.55 &	87.2\% \\
  Beethoven2Chopin &	6.19 &	95.3\% \\
  Bcp2Cjv &	6.07 &	95.7\% \\
  \hline
 \end{tabular}
 \end{center}
 \caption{ FAD\&Accuracy for the tasks.}
 \label{tab:2}
\end{table}

The accuracy of style transfer between the audio produced by the specified task and the original audio is presented in \tabref{tab:2} along with the results of FAD calculations. The results indicate good performance in terms of the timbre transfer of instruments and the transfer of compositional styles. However, the performance is poor in terms of genre transfer, which is consistent with the results of the subjective evaluation. This is an area that requires improvement in future research.

\subsubsection{Comparison With Other Models}
Our model was compared against a number of baseline models, including CycleGAN \cite{brunner2018symbolic}, UNIT \cite{liu2017unsupervised}, musicVAE \cite{roberts2018hierarchical}, and autoencoder \cite{bank2020autoencoders}, in order to show the validity of the model described in this work. \tabref{tab:3} presents the outcomes.

Note that these baseline models for style transfer are all one-to-one mappings. In this work, the input transfers use the same music clip, and they are trained independently. Only the spectrogram form is considered for the intermediate representation of the music. The same model, GuideDiff, is used for the generation of the waveform.

\begin{table}[h]
 \begin{center}
 \begin{tabular}{l | l l l l l}
  \hline
  \multirow{2}{1em}{Model} &	\multicolumn{5}{c}{Task}\\
  \cline{2-6}
	& p2v	& p2g & j2p	 & j2c & B2C\\
 \hline
CycleGAN	& 3.98 &	3.96 &	4.17 &	4.12 &	4.0\\
UNIT &	3.7	 &3.75	 & 3.5 &	3.62 &	3.71\\
musicVAE &	3.86 &	3.91 &	3.7 &	3.68 &	3.89\\
autoencoder &	3.5 &	3.56 &	3.4 &	3.45 &	3.52\\
ours &	4.23 &	4.09 &	3.91 &	4.02 &	4.07\\
  \hline
 \end{tabular}
\end{center}
 \caption{ MOS with the baseline comparison model.}
 \label{tab:3}
\end{table}

The comparison of the baseline models indicates that CycleGAN performs the best in terms of genre migration, which may be related to the fact that cycle consistency loss is taken into account in its direct matching of target domains at the feature level. However, our model achieved a result that is only about 0.1 points lower than the best. Additionally, our model outperforms the other baseline models in terms of the migration of musical instrument timbre and compositional style. Therefore, it can be concluded that the proposed model demonstrates superior performance in terms of flexible many-to-many musical style migration compared to the other baseline models.

\subsubsection{Evaluation of the audio generating model}
To demonstrate the performance and high-quality audio generation capabilities of GuideDiff, the proposed audio generation model, examples are presented in this section. Comparisons are made between the proposed model and WaveNet \cite{oord2016wavenet}, WaveRNN \cite{gritsenko2020spectral}, and WaveGAN \cite{yamamoto2020parallel}. All models use the same training set and are tested using the same spectrograms to generate audio. Both subjective and objective evaluation techniques are used to assess the quality of the generated audio. Testers will rate the audio quality on a scale of 1 to 5 for subjective evaluation. The results are presented in \tabref{tab:4}.

\begin{table}[h]
 \begin{center}
 \begin{tabular}{l  c c }
  \hline
  Model &	MOS(↑) &	IS(↑)\\
  \hline
WaveNet &	3.02 &	2.84\\
WaveGAN &	3.82 &	4.53\\
WaveRNN	 & 4.40 &	5.38\\
GuideDiff &	4.41 &	5.40\\

  \hline
 \end{tabular}
\end{center}
 \caption{ Comparison of audio generation models.}
 \label{tab:4}
\end{table}

The comparison demonstrates that our model performs similarly to the autoregressive model WaveRNN and surpasses the other baseline models. This suggests that the proposed model has excellent performance in producing high-quality audio.

\section{Conclusions}

In this work, we have designed an efficient DM-based framework for music style transfer. A latent layer was introduced into the framework, which effectively reduces the dimensionality of the data. A cross-attention mechanism is added to the latent layer. The transfer of styles is achieved by adding seed conditions to guide and complete the generation of transformations in the target domain. As for the generation of audio, this study proposes GuideDiff, a DM-based method for generating waveform audio. The method compresses the spectrogram into latent space via an encoder and transfers it to the U-Net. The potential signal is then decoded back into the waveform using diffusion guidance.

The experimental results demonstrate that the proposed model can achieve many-to-many style migration and generate high-quality music in comparison to previous approaches. Additionally, the model is capable of performing style migration and generating high-quality audio in real-time on a consumer-grade GPU. Given the excellent performance of this model, future work will utilize it to explore text-generated music.

\bibliography{icmc2023template}

\end{document}